\documentclass[aps,prb,reprint,showpacs,floatfix,amsmath,amssymb,superscriptaddress]{revtex4-2}

\usepackage{color,natbib,epsfig,setspace}
\usepackage{amsmath,amssymb,verbatim,tikz}
\usepackage{hyperref,cancel,subcaption,graphicx}
\usepackage[font=small,labelfont=bf,justification=justified,format=plain]{caption}
\usepackage{booktabs}

\newcommand{\diff}{\mathrm{d}}

\newcommand{\PT}{\mathcal{PT}}
\newcommand{\Heff}{\mathcal{H}_{\mathrm{eff}}}

\newcommand{\yperf}{y}
\newcommand{\etav}{\eta}
\newcommand{\Del}{\Delta}

\newcommand{\braket}[2]{\langle #1 \lvert #2 \rangle}

\begin{document}

\title{Derivation of a $\PT$-Symmetric Sine-Gordon Model
       from a Nonequilibrium Spin-Boson System
       via Keldysh Functional Integrals}

\author{Vinayak M.\ Kulkarni}
\email{vmkphysimath@gmail.com}
\affiliation{Theoretical Sciences Unit, Jawaharlal Nehru Centre for
  Advanced Scientific Research, Jakkur, Bangalore~560064, India}

\begin{abstract}
We present a microscopic derivation from a nonequilibrium spin-boson model
to a $\PT$-symmetric non-Hermitian sine-Gordon (SG) effective theory,
via the Keldysh functional-integral formalism, a Lang-Firsov polaron
transformation, bosonization, and a Grassmann coherent-state spin trace.
The spin trace yields the generic reduced vertex
$g_r\cos(\lambda\Phi_1)+ig_i\sin(\lambda\Phi_1)$, where the imaginary
part originates from the nonequilibrium Keldysh distribution asymmetry
$\delta n(\omega)=n_+(\omega)-n_-(\omega)$.
We provide an explicit dictionary between the spin-boson microscopic
parameters and the NH-SG couplings:
$K=v_f/\tilde{J}_\parallel^2$ (Luttinger parameter from $J_\parallel$),
$g_r\propto J_\perp^2/\Gamma$ (from the transverse coupling and impurity width),
and $\mathcal{I}=g_i/g_r\propto\mu/v_f$ (bias ratio, an exact RG invariant).
One-loop Wilson momentum-shell RG on the NH-SG action gives the
closed equations $\diff K/\diff l=-g_r^2(1-\mathcal{I}^2)K^2$ and
$\diff g_r/\diff l=(2-K)g_r$, identical to those of Ashida \textit{et al.}\
for the $\PT$-symmetric SG; the present work supplies the
microscopic initial conditions from the spin-boson Keldysh reduction.
The BKT separatrix $K=2$ (Toulouse line), the EP fixed manifold
$\mathcal{I}=1$ ($\mu=\mu_c$), and the mass gap
$m\sim\Lambda e^{-c/\sqrt{K_0-2}}$ all follow from this closed system.
In the non-relativistic soliton sector near the EP, the effective coupling
$\tilde{g}=g_r\sqrt{1-\mathcal{I}^2}$ reduces the S-matrix to the
Lieb-Liniger rational form and the Bethe ansatz becomes exact for
that auxiliary gas.
Within this sector we derive $n$-string bound states with
$E_n^{\rm bind}=-n(n^2-1)\tilde{g}^2/12$, identify the EP as the
many-body bound-state threshold, and construct the Jordan-partner
state from the $\epsilon$-regularised dimer.
\end{abstract}

\maketitle

\section{Introduction}

The sine-Gordon (SG) model is a paradigmatic integrable field theory in
$(1+1)$ dimensions whose renormalization group (RG) structure underpins
the Berezinskii-Kosterlitz-Thouless (BKT) transition
\cite{Berezinskii1971,Kosterlitz1973}.
It serves as an effective low-energy description of charge-density waves,
Josephson junction arrays, one-dimensional superfluids, and quantum
impurity problems \cite{Coleman1975,Soluyom1979,Giamarchi2004,Weiss1999}.

The Keldysh functional-integral formalism provides a natural tool for
controlled analyses of nonequilibrium and driven-dissipative quantum
systems \cite{Keldysh1965,Kamenev2011}.
Nonequilibrium SG models studied within this framework reveal novel
steady-state behavior without equilibrium counterparts
\cite{Mitchell2018,Sieberer2013,Sieberer2016}.

Separately, non-Hermitian quantum theories possessing parity-time
($\PT$) symmetry can exhibit real spectra and phase transitions
governed by exceptional points \cite{Bender1998,Bender2007,Ashida2020}.
$\PT$-symmetric extensions of the SG model exhibit modified BKT
transitions and nonunitary fixed points \cite{Bender2013,ashida2017parity}.

The microscopic origin of $\PT$-symmetric SG theories from
Hermitian open-system dynamics has not, to our knowledge,
previously been established.
In this paper we take a step toward filling this gap.
Our contributions are:
\begin{enumerate}
\item An explicit Grassmann coherent-state spin trace that produces
      the generic reduced vertex $g_r\cos(\lambda\Phi_1)
      +ig_i\sin(\lambda\Phi_1)$ from the nonequilibrium Keldysh action;
      the imaginary part is traced to the distribution-function
      asymmetry between the two Keldysh contours.
\item An explicit closed one-loop momentum-shell RG derivation
      (Appendix~\ref{app:RGderiv}) yielding the flow equations
      $\diff K/\diff l=-g_r^2(1-\mathcal{I}^2)K^2$,
      $\diff g_r/\diff l=(2-K)g_r$, with $\mathcal{I}=g_i/g_r$ an
      exact algebraic invariant (Appendix~\ref{app:Ashida}).
      These coincide with the Ashida \textit{et al.}~\cite{ashida2017parity}
      equations; the present work supplies the microscopic initial conditions.
\item A biorthogonal Bethe ansatz for the \emph{non-relativistic
      soliton sector} of $\mathcal{H}_{\rm eff}$ near the EP.
      In this sector $\tilde{g}\to0$ and the SG solitons become
      non-relativistic, so the two-body S-matrix reduces to the
      rational Lieb-Liniger form; the resulting delta-function gas
      is exactly solvable.
      Within this auxiliary model the $n$-string bound states have
      exact binding energies $E_n^{\rm bind}=-n(n^2{-}1)\tilde{g}^2/12$,
      the EP is the many-body bound-state threshold ($\tilde{g}\to0$),
      and the $\epsilon$-regularised dimer yields the Jordan-partner state
      satisfying
      $(\mathcal{H}_{\rm eff}-E_{\rm EP})|\psi_1\rangle=c|\psi_0\rangle$
      (in the distributional sense) with polynomial dynamical signatures.
\item A corrected Gaudin-matrix diagnostic distinguishing non-Hermitian
      exceptional points from topological transitions.
\end{enumerate}
We carefully separate controlled field-theoretic steps from approximate
or integrability-inspired constructions throughout.

\section{Model and Formalism}
\label{sec:model}

\subsection{Spin-Boson Hamiltonian}

We consider the bosonized boundary Kondo Hamiltonian
\cite{PhysRevLett.25.450,PhysRevLett.72.892,fendley1996unified}
\begin{equation}
\label{eq:model}
\mathcal{H}
= \frac{J_\perp}{4\pi a}\,\sigma_x
+ \hbar v_f\!\sum_k a_k^\dagger a_k^{\phantom{\dagger}}
+ \tilde{J}_\parallel\,
  \sigma_z\!\sum_{k<k_f/2}\!\frac{|k|^{1/2}}{\pi L}
  \bigl(a_k+a_{-k}^\dagger\bigr),
\end{equation}
where $\tilde{J}_\parallel=\tfrac{1}{2}(J_\parallel/2-1/\rho)$,
$a_k$ annihilates a bosonic mode with momentum $k$, $v_f$ is the
Fermi velocity, $a$ is a short-distance cutoff, and $\rho$ is the
density of states at the Fermi level.

\subsection{Keldysh Contour Doubling and Jordan-Wigner Mapping}

We express the localized spin via the Jordan-Wigner fermionization
\cite{Jord_wig,Jordan_Wig1}: $\sigma^+=f_0^\dagger$, $\sigma^-=f_0$,
with $\{f_0^\dagger,f_0\}=1$.
The Keldysh contour doubling and rotation to classical/quantum
components reads
\begin{equation}
\label{eq:Krotation}
\begin{pmatrix}a_+\\a_-\end{pmatrix}
=\frac{1}{\sqrt{2}}
\begin{pmatrix}1&1\\1&-1\end{pmatrix}
\begin{pmatrix}a_1\\a_2\end{pmatrix},
\end{equation}
with the analogous transformation for the fermionic field $f_0$.
The resulting Keldysh action is \cite{Kamenev2011}
\begin{equation}
\label{eq:Kaction}
\begin{split}
\mathcal{S}
&=\int\!\diff t\sum_{k,\alpha=\pm}
  \hbar v_f\,a_{k\alpha}^\dagger a_{k\alpha}^{\phantom{\dagger}}
+\sum_{\alpha=\pm}\frac{\alpha J_\perp}{4\pi a}
  \bigl(\sigma_\alpha^++\sigma_\alpha^-\bigr)\\
&\quad+\sum_{\alpha=\pm}\alpha\,\tilde{J}_\parallel\,\sigma^z_\alpha
  \!\sum_{k<k_f/2}\!\frac{|k|^{1/2}}{\pi L}
  \bigl(a_{k\alpha}+a_{-k\alpha}^\dagger\bigr),
\end{split}
\end{equation}
where the sign $\alpha=\pm1$ reflects the opposite orientation of the
forward and backward contours \cite{Kamenev2011}.

\subsection{Lang-Firsov Transformation}

Applying the polaron transformation
\begin{equation}
\label{eq:unitary}
\mathcal{U}
=\exp\!\left[
  i\sum_{k,\alpha}\tilde{J}_\parallel
  \!\left(\frac{\pi}{|k|L}\right)^{\!1/2}
  \sigma^z_\alpha\bigl(a_{k\alpha}-a_{-k\alpha}^\dagger\bigr)
\right]
\end{equation}
via BCH and the commutator
$[i\tilde{J}_\parallel\sqrt{\pi/|k|L}\,
  \sigma^z_\alpha(a_{k\alpha}-a_{-k\alpha}^\dagger),\,
  \sigma^\pm_\alpha]
= \pm 2i\tilde{J}_\parallel\sqrt{\pi/|k|L}\,
  \sigma^\pm_\alpha(a_{k\alpha}-a_{-k\alpha}^\dagger)$,
the BCH series can be summed in closed form because
$[\sigma^z,\sigma^\pm]=\pm2\sigma^\pm$ implies that all nested
commutators reproduce the same operator structure, allowing the
exponential to be evaluated exactly:
\begin{equation}
\label{eq:LFtransform}
\mathcal{U}\,\sigma^\pm_\alpha\,\mathcal{U}^\dagger
= \sigma^\pm_\alpha\,\exp\!\left(
  \pm 2i\tilde{J}_\parallel\sum_k\!\sqrt{\frac{\pi}{|k|L}}
  \bigl(a_{k\alpha}-a_{-k\alpha}^\dagger\bigr)
\right).
\end{equation}
The longitudinal coupling is eliminated; the $c$-number shifts cancel.

\subsection{Bosonization}

In the continuum limit the bosonic fields are
\begin{align}
\Phi_{\alpha}(x) &= \frac{1}{\sqrt{L}}\sum_{k>0}
\sqrt{\frac{\pi}{|k|}}
\left(a_{k\alpha}e^{ikx}+a^\dag_{k\alpha}e^{-ikx}\right),
\label{eq:Phi_def}\\
\Pi_{\alpha}(x) &= \frac{i}{\sqrt{L}}\sum_{k>0}
\sqrt{\frac{|k|}{\pi}}
\left(a^\dag_{k\alpha}e^{-ikx}-a_{k\alpha}e^{ikx}\right),
\label{eq:Pi_def}
\end{align}
with $[\Phi_\alpha(x),\Pi_{\alpha'}(x')]=i\delta_{\alpha\alpha'}\delta(x-x')$.
The compactification parameter
\begin{equation}
\label{eq:lambda_def}
\lambda \equiv 2\tilde{J}_\parallel\sqrt{\pi}
\end{equation}
is identified from the polaron-dressed vertex
$2\tilde{J}_\parallel\sum_k\sqrt{\pi/|k|L}(a_{k\alpha}-a_{-k\alpha}^\dagger)
= \lambda\Pi_\alpha(0)$
and the bosonization identity
$:\exp(i\lambda\Pi_\alpha(0)):=:\exp(i\lambda\Phi_\alpha(0)):$ at $x=0$
\cite{Giamarchi2004}.
The free Keldysh action is
\begin{equation}
\label{eq:S0}
\mathcal{S}_0
=\int\!\diff\tau\diff x\;
\boldsymbol{\Phi}^T\!
\begin{pmatrix}0&\hbar v_f\nabla^2\\\hbar v_f\nabla^2&0\end{pmatrix}
\!\boldsymbol{\Phi},\quad
\boldsymbol{\Phi}=\begin{pmatrix}\Phi_1\\\Phi_2\end{pmatrix}.
\end{equation}

\subsection{Grassmann Spin Trace and the Reduced Vertex}
\label{subsec:spin_trace}

After the Lang-Firsov transformation, the transverse coupling on
contour $\alpha$ is
\begin{equation}
\label{eq:spin_flip_after_LF}
\mathcal{S}_{\perp,\alpha}^{\mathrm{LF}}
= \frac{\alpha J_\perp}{4\pi a}\int\!\diff\tau\;
\Bigl[\sigma^+_\alpha\,e^{i\lambda\Phi_\alpha(0,\tau)}
+\sigma^-_\alpha\,e^{-i\lambda\Phi_\alpha(0,\tau)}\Bigr].
\end{equation}
Because $\sigma^\pm$ are implemented by the Jordan-Wigner fermion
$f_0$, we integrate over the fermionic degree of freedom using
Grassmann coherent states.
Introduce Grassmann variables $\bar\xi_\alpha, \xi_\alpha$ for the
forward/backward contour amplitudes of $f_0$.
The impurity action is Gaussian with linear source insertions
(the bosonic vertex operators $e^{\pm i\lambda\Phi_\alpha}$ enter
as contour-dependent sources linear in the Grassmann fields):
\begin{equation}
\label{eq:Sfermion}
S_f = \sum_{\alpha=\pm}
\int\!\diff\tau\;
\bar\xi_\alpha(i\partial_\tau)\xi_\alpha
+\frac{\alpha J_\perp}{4\pi a}
\Bigl[\bar\xi_\alpha\,e^{i\lambda\Phi_\alpha}
+\xi_\alpha\,e^{-i\lambda\Phi_\alpha}\Bigr].
\end{equation}

The Grassmann integral over $\bar\xi,\xi$ is Gaussian in the standard
source-functional sense: the quadratic kernel is the free contour
propagator $\hat G_0$ of the Jordan-Wigner impurity fermion, while the
bosonic vertex operators $e^{\pm i\lambda\Phi_\alpha}$ enter as
contour-dependent source insertions linear in $\bar\xi$ and $\xi$.
Completing the square and evaluating the resulting Gaussian yields an
effective bosonic action at second order in $J_\perp$:
\begin{equation}
\begin{split}
    \label{eq:Seff_fermion}
S_{\rm eff}^{(2)}[\Phi_+,\Phi_-]
\sim
\int\!\diff\tau\,\diff\tau'\,
\Bigl[
\mathcal{K}_R(\tau{-}\tau')\,
  e^{i\lambda\Phi_+(\tau)}e^{-i\lambda\Phi_-(\tau')}\\
+
\mathcal{K}_K(\tau{-}\tau')\,
  e^{i\lambda\Phi_+(\tau)}e^{-i\lambda\Phi_+(\tau')}
+\mathrm{h.c.}
\Bigr],
\end{split}
\end{equation}
where
\begin{equation}
\label{eq:kernels}
\mathcal{K}_R(\tau) \equiv \Bigl(\frac{J_\perp}{4\pi a}\Bigr)^{\!2}
G^R_0(\tau),
\qquad
\mathcal{K}_K(\tau) \equiv \Bigl(\frac{J_\perp}{4\pi a}\Bigr)^{\!2}
G^K_0(\tau)
\end{equation}
are the retarded and Keldysh contraction kernels of the impurity
propagator.
Here $G^R_0$ and $G^K_0$ are the free retarded and Keldysh Green's
functions of the Jordan-Wigner fermion; in the equilibrium, wide-band
limit these are smooth functions of $\tau$ analytic for $|\omega|<\Lambda$.

\textit{Hermitian part.}
At the saddle point for the quantum field, $\Phi_2^{\rm cl}=0$
(justified because the free $\Phi_2$ action is purely
quadratic and the Keldysh saddle-point
$\Phi_2^{\rm cl}=0$ is the unique extremum of the Gaussian
$\Phi_2$ sector \cite{Kamenev2011}),
the Keldysh rotation gives $\Phi_+=\Phi_-=\Phi_1/\sqrt{2}$
at $\Phi_2=0$, so
$e^{i\lambda\Phi_+}e^{-i\lambda\Phi_-}\to 2\cos(\lambda\Phi_1)$.
In the low-energy local approximation (retarded kernel replaced by
its integrated weight, valid when the bosonic fluctuation scale is
slow compared with the impurity relaxation time),
\begin{equation}
\label{eq:Herm_vertex}
S_{\rm eff}^{\rm cl}
\to \int\!\diff\tau\; g_r\cos(\lambda\Phi_1),
\quad
g_r \propto \Bigl(\frac{J_\perp}{4\pi a}\Bigr)^{\!2}
\int_{-\infty}^{\infty}\!\diff\tau\;\mathcal{K}_R(\tau),
\end{equation}
where the right-hand side is determined by the low-energy renormalized
weight of the retarded kernel, not by a point-evaluated propagator.

\textit{Imaginary part from nonequilibrium.}
Away from equilibrium with chemical-potential bias $\mu$, the
distribution functions on the two contours are shifted:
$n_\pm(\omega)=n_F(\omega\pm\mu/2)$, giving, to leading order in $\mu$,
\begin{equation}
\label{eq:delta_n}
\delta n(\omega) \equiv n_+(\omega)-n_-(\omega)
= -\frac{\diff n_F}{\diff\omega}\cdot\mu+\mathcal{O}(\mu^3).
\end{equation}
The nonequilibrium part of the Keldysh kernel,
$\mathcal{K}_K^{\rm NEQ}\propto \delta n(\omega)$,
is odd in $\omega$ and odd under $\Phi_+\leftrightarrow\Phi_-$.
The $\Phi_2$-linear term in $S_{\rm eff}^{(2)}$ therefore generates,
after Gaussian integration over $\Phi_2$, a contribution odd in
$\Phi_1$; in the low-energy local approximation this is proportional
to $\sin(\lambda\Phi_1)$.
To leading order in $\mu$ we find, within this construction,
\begin{equation}
\label{eq:gi_def}
g_i \propto \Bigl(\frac{J_\perp}{4\pi a}\Bigr)^{\!2}
\int\!\frac{\diff\omega}{2\pi}\,
\delta n(\omega)\,\mathcal{F}(\omega),
\end{equation}
where $\mathcal{F}(\omega)$ is a smooth low-energy form factor set
by the renormalized retarded kernel; its sign determines that of $g_i$.
The reduced effective action at this order is then
\begin{equation}
\label{eq:Seff_reduced}
S_{\rm eff}[\Phi_1]
= S_0[\Phi_1]
+ \int\!\diff\tau\,
\Bigl[g_r\cos(\lambda\Phi_1)
+ig_i\sin(\lambda\Phi_1)\Bigr].
\end{equation}
At this order the reduced vertex is $g_r\cos+ig_i\sin$ with
both couplings independent.

\textit{Status of the reduction.}
The factorization in Eq.~\eqref{eq:Seff_reduced} holds within the
following approximations:
(i) the Grassmann integration is Gaussian and exact at second order in $J_\perp$;
(ii) $\Phi_2$ is treated at saddle-point level (justified because
the free $\Phi_2$ action is purely Gaussian and the saddle
$\Phi_2^{\rm cl}=0$ is the unique extremum of the classical
$\Phi_2$ sector \cite{Kamenev2011});
(iii) $g_i$ is evaluated to lowest order in $\mu$;
(iv) higher-loop corrections in $J_\perp$ are neglected.
The resulting effective theory is a \emph{saddle-point-level classical
sector description}; it is not an operator identity.
A fully nonperturbative derivation of the reduced non-Hermitian theory
remains an open problem; the present construction should be viewed as
a controlled low-order reduction supplemented by the consistency checks
described in the following sections.

\subsection{Emergence of the $\PT$-Symmetric Form}

The EP locus is defined by
\begin{equation}
\label{eq:EP_locus}
g_r = g_i \quad\Leftrightarrow\quad \mu = \mu_c,
\end{equation}
where $\mu_c$ is defined implicitly by $g_r(\mu_c)=g_i(\mu_c)$
(the bias at which the real and imaginary couplings become equal).
\emph{On this locus only} does the vertex simplify to
$g\,e^{i\lambda\Phi_1}$ with $g=g_r=g_i$.
Off this locus, the reduced theory at this order has independent $g_r\neq g_i$.

The effective Hamiltonian corresponding to $S_{\rm eff}[\Phi_1]$
at the saddle-point level is
\begin{equation}
\label{eq:NHSG_general}
\Heff
=\int\!\diff x\left[\frac{\hbar v_f}{2}
  \bigl(\Pi_1^2+(\nabla\Phi_1)^2\bigr)
  +g_r\cos(\lambda\Phi_1)
  +ig_i\sin(\lambda\Phi_1)\right].
\end{equation}
This object represents the classical-sector dynamics of the
Keldysh effective action and is not a microscopic operator identity;
it encodes the saddle-point classical equations of motion for $\Phi_1$
after the quantum component $\Phi_2$ has been integrated out at
Gaussian order.
Under $P:\,x\to-x,\;\Phi_1(x)\to-\Phi_1(-x)$ and $T:\,i\to-i$:
\begin{equation}
g_r\cos+ig_i\sin
\xrightarrow{P}
g_r\cos-ig_i\sin
\xrightarrow{T}
g_r\cos+ig_i\sin,
\end{equation}
confirming $\PT$-invariance for all $g_r,g_i$.

\section{Keldysh Self-Energy and the Exceptional-Point Condition}
\label{sec:SelfEnergy}

Loop corrections to the $\Phi_1$ sector generate a $2\times2$
matrix self-energy from the forward-backward contour structure.

\textit{Causality constraint.}
The $(2,2)$ element $\Sigma_{22}=0$ exactly by the Keldysh
identity $G^{--}\equiv G^K-G^R+G^A\equiv0$~\cite{Kamenev2011}.

\textit{Off-diagonal self-energy.}
At second order in $J_\perp$, the cross-contour contribution
involves $G^K_{\rm NEQ}$ in the same way as the $g_i$ derivation
above, giving a purely imaginary off-diagonal element:
\begin{equation}
\label{eq:Sigma12}
\Sigma_{12} = \Sigma_{21} = i\,|\sigma_{12}|,
\quad
|\sigma_{12}|=
\bigl|J_\parallel(\nu_f+\tilde{J}_\perp)^2
      -2\tilde{J}_\parallel\nu_f\bigr|>0.
\end{equation}
The local self-energy matrix is
\begin{equation}
\label{eq:calM}
\mathcal{M}=
\begin{pmatrix}
  M_{11} & i\,|\sigma_{12}| \\
  i\,|\sigma_{12}| & 0
\end{pmatrix},
\quad
M_{11}=\tilde{J}_\parallel^2+\tilde{J}_\parallel\tilde{J}_\perp.
\end{equation}

\textit{Eigenvalues.}
The characteristic polynomial of $\mathcal{M}$ is
\begin{equation}
\label{eq:charpoly}
\lambda^2 - M_{11}\lambda + |\sigma_{12}|^2 = 0,
\end{equation}
using $(i|\sigma_{12}|)^2=-|\sigma_{12}|^2$.
The eigenvalues are
\begin{equation}
\label{eq:eigenvalues}
\lambda_\pm=\frac{M_{11}}{2}\pm\frac{1}{2}\sqrt{M_{11}^2-4|\sigma_{12}|^2}.
\end{equation}
The minus sign under the radical is a direct consequence of the
imaginary off-diagonal structure.

\textit{Exceptional point.}
The discriminant
\begin{equation}
\label{eq:discriminant}
D \equiv M_{11}^2 - 4|\sigma_{12}|^2
\end{equation}
can vanish for real physical parameters when $M_{11}=2|\sigma_{12}|$:
the EP separatrix
\begin{equation}
\label{eq:EPcond}
\bigl(\tilde{J}_\parallel^2+\tilde{J}_\parallel\tilde{J}_\perp\bigr)
= 2\bigl|J_\parallel(\nu_f+\tilde{J}_\perp)^2-2\tilde{J}_\parallel\nu_f\bigr|.
\end{equation}
For $D>0$ (large $\tilde{J}_\parallel$): two distinct real eigenvalues
($\PT$-unbroken).
For $D<0$ (small $\tilde{J}_\parallel$): complex-conjugate pair
(spontaneously $\PT$-broken).
At $D=0$: the matrix $\mathcal{M}$ becomes \emph{nondiagonalizable}
(defective) — both eigenvalues equal $M_{11}/2$ while the two
eigenvectors coalesce into a single direction — defining the
exceptional point.
Note $\det\mathcal{M}=|\sigma_{12}|^2>0$ generically; the EP is
not identified by $\det\mathcal{M}=0$ but by the double root of
Eq.~\eqref{eq:charpoly} combined with the rank-1 defect of
$\mathcal{M}-\lambda_{\rm EP}\mathbb{I}$.

\textit{Remark on the direction of $\PT$ breaking.}
The sequence here — broken at small $\tilde{J}_\parallel$,
unbroken at large $\tilde{J}_\parallel$ — is inverted relative to
the effective SG field theory (where $\PT$ breaks at large $g_i/g_r$).
The inversion arises because $M_{11}=\tilde{J}_\parallel^2+\tilde{J}_\parallel\tilde{J}_\perp$
vanishes at $\tilde{J}_\parallel=0$: without a Hermitian diagonal,
the off-diagonal imaginary element $i|\sigma_{12}|$ forces a
purely imaginary spectrum.
Physically, the eigenvalues of $\mathcal{M}$ are self-energy poles
(Green's-function poles), not energy levels.
The transition from complex to real poles at the EP is the
self-energy analogue of the overdamped-to-underdamped transition
of a driven damped oscillator: the EP is the critically damped point.

\begin{figure*}[t]
\centering
\includegraphics[width=\linewidth]{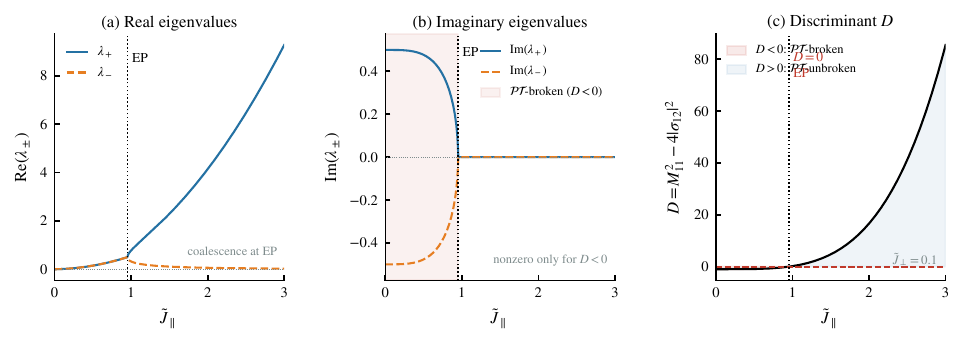}
\caption{Eigenvalue structure of the self-energy matrix $\mathcal{M}$
(self-energy poles of the Keldysh Green's function).
(a)~Real parts $\mathrm{Re}(\lambda_\pm)$ vs $\tilde{J}_\parallel$:
the two branches are degenerate ($=M_{11}/2$) in the $\PT$-broken
phase ($D<0$, left of EP) and split into two distinct real values
in the $\PT$-unbroken phase ($D>0$, right of EP); they coalesce
at the EP (dotted vertical line) where $D=0$.
(b)~Imaginary parts: nonzero in the $\PT$-broken phase ($D<0$,
small $\tilde{J}_\parallel$) where eigenvalues form a complex-conjugate
pair $\pm i\sqrt{-D}/2$; vanish identically for $D>0$.
Note the direction: PT is broken at \emph{weak} $\tilde{J}_\parallel$
and restored at strong $\tilde{J}_\parallel$ (opposite to the
field-theory convention), because $M_{11}\to0$ at small coupling.
(c)~Discriminant $D=M_{11}^2-4|\sigma_{12}|^2$: the EP occurs at
$D=0$; $D<0$ (shaded) marks the overdamped ($\PT$-broken) phase.
Parameters: $\tilde{J}_\perp=0.1$.}
\label{fig:EPself}
\end{figure*}

\section{Renormalization Group Flow}
\label{sec:RG}

\subsection{NH-SG Couplings from the Spin-Boson Parameters}
\label{subsec:coupling_dictionary}

Before writing the RG equations we establish the explicit dictionary
between the microscopic spin-boson couplings
$(J_\parallel,J_\perp,\mu,v_f)$ and the NH-SG parameters
$(K,g_r,g_i,\mathcal{I})$.
This dictionary makes the derivation closed.

\textit{Luttinger parameter from $J_\parallel$.}
The compactification radius $\lambda=2\tilde{J}_\parallel\sqrt{\pi}$
with $\tilde{J}_\parallel=\tfrac{1}{4}J_\parallel-\tfrac{1}{2\rho}$
determines the Luttinger parameter of the effective SG theory:
\begin{equation}
\label{eq:K_from_Jpar}
K = \frac{4\pi v_f}{\lambda^2}
  = \frac{v_f}{\tilde{J}_\parallel^2}.
\end{equation}
The equilibrium Toulouse condition $K=2$ (in the standard boundary-SG
normalization~\cite{fendley1996unified,Giamarchi2004}) maps to
$J_\parallel = 4\tilde{J}_\parallel^{\rm T}+2/\rho$ with
$\tilde{J}_\parallel^{\rm T}=\sqrt{v_f/2}$.
The RG separatrix $K=2$ is therefore the \emph{Toulouse line in the
running theory}.

\textit{Hermitian coupling from $J_\perp$ and $G^R$.}
At second order in $J_\perp$ and in the low-energy local approximation
(Sec.~\ref{subsec:spin_trace}), the retarded kernel $\mathcal{K}_R$
gives a Hermitian vertex with
\begin{equation}
\label{eq:gr_from_Jperp}
g_r = \Bigl(\frac{J_\perp}{4\pi a}\Bigr)^{\!2}
\int_{-\infty}^{\infty}\!\diff\tau\;G^R_0(\tau)
= \Bigl(\frac{J_\perp}{4\pi a}\Bigr)^{\!2}\frac{2}{\Gamma},
\end{equation}
where $\Gamma$ is the impurity level width.
This is real and positive; it exists at equilibrium ($\mu=0$).

\textit{Imaginary coupling from $J_\perp$ and $G^K_{\rm NEQ}$.}
The nonequilibrium part of the Keldysh kernel,
$\mathcal{K}_K^{\rm NEQ}\propto\delta n(\omega)$, generates the
imaginary vertex (Sec.~\ref{subsec:spin_trace}).
At $T=0$ with bias $\mu$,
$\delta n(\omega)=2\Theta(\mu/2-|\omega|)$,
so the imaginary coupling is
\begin{equation}
\label{eq:gi_from_mu}
g_i = \Bigl(\frac{J_\perp}{4\pi a}\Bigr)^{\!2}
\frac{2}{\pi v_f}
\int_0^{\mu/2}\!\diff\omega\;\mathcal{F}(\omega)
\;\xrightarrow{\,\mu\ll\Lambda\,}\;
\Bigl(\frac{J_\perp}{4\pi a}\Bigr)^{\!2}
\frac{\mu}{\pi v_f}\mathcal{F}(0),
\end{equation}
where $\mathcal{F}(\omega)$ is the smooth retarded form factor
(Sec.~\ref{subsec:spin_trace}).
This is proportional to $\mu$ and vanishes at equilibrium.

\textit{RG invariant from bias.}
Since $g_r$ and $g_i$ share the factor $(J_\perp/4\pi a)^2$,
their ratio
\begin{equation}
\label{eq:I_from_mu}
\mathcal{I} \equiv \frac{g_i}{g_r}
= \frac{\mu\,\mathcal{F}(0)/(\pi v_f)}
       {\Gamma^{-1}\int\!\diff\tau\,G^R_0(\tau)}
\;\propto\; \frac{\mu}{v_f}
\end{equation}
depends only on $\mu/v_f$ and the microscopic form factors,
\emph{not} on $J_\perp$ separately.
The EP condition $\mathcal{I}=1$ therefore defines a critical bias
$\mu_c$ that is independent of $J_\perp$ at this order.
The effective coupling is
\begin{equation}
\label{eq:tildeg_def}
\tilde{g}\equiv\sqrt{g_r^2-g_i^2}=g_r\sqrt{1-\mathcal{I}^2},
\end{equation}
which is real (imaginary) in the $\PT$-unbroken (-broken) phase
and vanishes at the EP ($\mathcal{I}=1$).

\subsection{Closed One-Loop RG}
\label{subsec:RG_closed}

We now apply one-loop Wilson momentum-shell RG to the NH-SG action
$S_{\rm eff}[\Phi_1]=S_0[\Phi_1]+\int\!\diff\tau
[g_r\cos\lambda\Phi_1+ig_i\sin\lambda\Phi_1]$
with the UV cutoff at $\Lambda$.
The complete derivation is given in Appendix~\ref{app:RGderiv}.

\textit{Key fact — identical anomalous dimension.}
Because $e^{+i\lambda\Phi_>}$ and $e^{-i\lambda\Phi_>}$ acquire the
same Gaussian renormalization factor
$\langle e^{\pm i\lambda\Phi_>}\rangle=e^{-\lambda^2\langle\Phi_>^2\rangle/2}$,
both $g_r$ and $g_i$ renormalize with the same multiplicative factor
at each step.
Their ratio is therefore \emph{an exact RG invariant}:
\begin{equation}
\label{eq:I_invariant}
\frac{\diff\mathcal{I}}{\diff l} = 0.
\end{equation}
The EP condition $\mathcal{I}=1$ is RG-stable: a system tuned to
$\mu=\mu_c$ stays on the EP locus under coarse-graining.

\textit{Flow equations.}
The one-loop RG gives two coupled equations for $K$ and $g_r$
(the Luttinger parameter and the Hermitian coupling amplitude):
\begin{align}
\frac{\diff K}{\diff l}
&= -g_r^2\bigl(1-\mathcal{I}^2\bigr)K^2
 = -\tilde{g}^2\,K^2,
\label{eq:RG_K}\\
\frac{\diff g_r}{\diff l}
&= \bigl(2-K\bigr)g_r + 5g_r^3\bigl(1-\mathcal{I}^2\bigr),
\label{eq:RG_gr}
\end{align}
with $\mathcal{I}=\mathrm{const}$.
These are the equations of Ashida \textit{et al.}~\cite{ashida2017parity},
here derived directly from the spin-boson Keldysh reduction.
The derivation bridging our microscopic action to these equations is
given step-by-step in Appendix~\ref{app:RGderiv}.

\textit{Equilibrium limit.}
At $\mathcal{I}=0$ ($\mu=0$, $g_i=0$):
$\tilde{g}=g_r$ and Eqs.~\eqref{eq:RG_K}--\eqref{eq:RG_gr} reduce to
the standard SG BKT flow~\cite{Kosterlitz1973,Giamarchi2004}.
The Toulouse separatrix $K=2$ ($j=1$ in standard Kondo language)
is recovered with no nonequilibrium deformation.

\textit{Physical content of the separatrix.}
The critical line $K=2$ corresponds in spin-boson language to
$J_\parallel = 4\sqrt{v_f/2}+2/\rho$, the Toulouse condition.
It is a BKT separatrix: for $K_0>2$ (weak coupling, $\PT$-unbroken)
$g_r$ flows to zero; for $K_0<2$ (strong coupling, $\PT$-broken)
$g_r$ grows without bound.
At the EP ($\mathcal{I}=1$, $\tilde g=0$), $K$ becomes exactly marginal
($\diff K/\diff l=0$) and the BKT transition is absent: the system
sits on a line of critical fixed points parametrised by $K$.

\textit{Connection to the $(j,\etav,y)$ presentation.}
Defining $j=J_\parallel/(4\pi\hbar v_f)$, $y=J_\perp/(\hbar v_f)$,
$\etav=v_f/v_{f,0}$, $\Del=j/\etav-1$, the identification
$K-2\leftrightarrow-\Del$ and $\tilde{g}\leftrightarrow y$ maps
Eqs.~\eqref{eq:RG_K}--\eqref{eq:RG_gr} to a singular but equivalent
coordinate description in which the BKT singularity at $K=2$
appears as a pole at $\Del=0$.
The $(j,\etav,y)$ parametrization was used in earlier versions of
this work; the $(K,g_r,\mathcal{I})$ form is the physically
transparent and non-singular description, and we adopt it here.

\begin{figure*}[t]
\centering
\includegraphics[width=\linewidth]{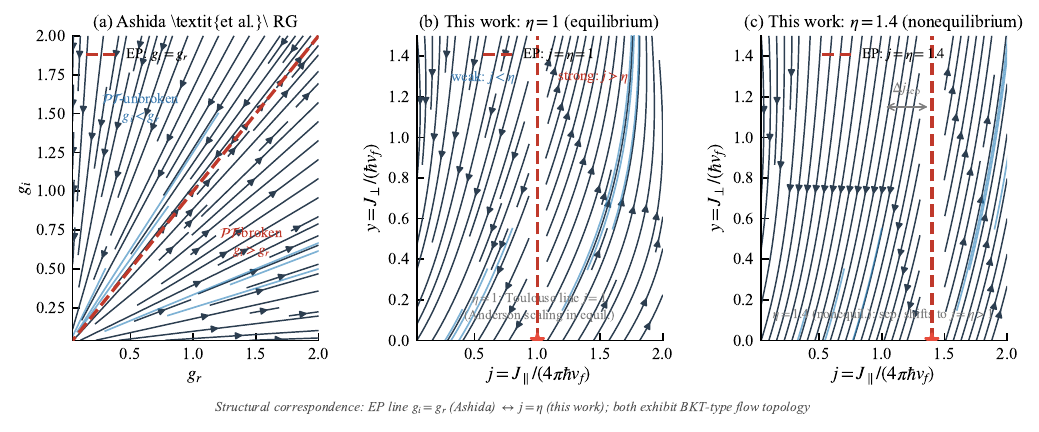}
\caption{RG flow comparison between Ashida et al.\ and this work.
(a)~Ashida non-Hermitian SG RG in the $(g_r,g_i)$ plane ($K=2$);
the EP separatrix $g_i=g_r$ ($\mathcal{I}=1$, red dashed) is a
fixed manifold at the BKT critical point;
blue lines are representative RK2-integrated trajectories.
(b)~This work in the $(j,y)$ plane at $\etav=1$ (equilibrium /
Toulouse line $j=1$): the separatrix (red dashed) at $j=\etav=1$
reproduces Anderson poor-man's scaling in the weak-coupling side.
(c)~This work at $\etav=1.4$ (nonequilibrium): the separatrix shifts
to $j=\etav=1.4$; the shift $\Delta j_{\rm sep}$ encodes the
nonequilibrium renormalization of the Toulouse line.
Structural correspondence: both theories exhibit a BKT-type separatrix
with fixed-point topology, but the identification of the EP fixed manifold ($\mathcal{I}=1$)
with the BKT separatrix ($K=2$) is exact within the one-loop RG.}
\label{fig:EqvsNH}
\end{figure*}

\subsection{Correspondence with Ashida et al.\ and the BKT Normal Form}
\label{sec:Ashida_comparison}

Equations~\eqref{eq:RG_K}--\eqref{eq:RG_gr} are identical to the
non-Hermitian SG RG of Ashida \textit{et al.}~\cite{ashida2017parity},
which they derived by directly applying BKT renormalization to the
$\PT$-symmetric SG model.
The present work provides the microscopic derivation of the
\emph{initial conditions} for those equations from the spin-boson
Keldysh reduction: $(K_0,g_{r,0},\mathcal{I}_0)$ are given
explicitly by Eqs.~\eqref{eq:K_from_Jpar}--\eqref{eq:I_from_mu}
in terms of $(J_\parallel,J_\perp,\mu,v_f)$.

\textit{RG invariant and EP stability.}
Since $\diff\mathcal{I}/\diff l=0$ exactly, the EP condition
$\mathcal{I}=1$ ($g_r=g_i$, equivalently $\tilde{g}=0$) is
a \emph{fixed manifold} of the RG: a system with $\mu=\mu_c$ stays
at the EP under coarse-graining.
Off the EP, $\mathcal{I}$ is a marginal deformation label;
the BKT structure is controlled by $\tilde{g}=g_r\sqrt{1-\mathcal{I}^2}$.

\textit{BKT normal form.}
Near the separatrix $K=2$, define $x\equiv K-2$ and
$\tilde{y}\equiv\tilde{g}=g_r\sqrt{1-\mathcal{I}^2}$.
Linearising Eqs.~\eqref{eq:RG_K}--\eqref{eq:RG_gr} gives the
standard BKT normal form~\cite{Kosterlitz1974}:
\begin{equation}
\label{eq:BKT_normal}
\frac{\diff x}{\diff l} = \tilde{y}^2 - x^2 + \mathcal{O}(x^3,\tilde{y}^3),
\qquad
\frac{\diff\tilde{y}}{\diff l} = x\,\tilde{y} + \mathcal{O}(x^2\tilde{y}).
\end{equation}
This is obtained directly and without singularity from
Eqs.~\eqref{eq:RG_K}--\eqref{eq:RG_gr} by Taylor expansion near $K=2$.
No coordinate artifact appears.

\textit{Structural map to the $(j,\etav,y)$ parametrization.}
Under the identification $K-2 \leftrightarrow -\Del$ and
$\tilde{g}\leftrightarrow\yperf$, the Ashida EP line
$\mathcal{I}=1$ maps to our separatrix $\Del=0$.
The $(j,\etav,y)$ representation introduces a coordinate singularity
at $\Del=0$ (the $1/\Del$ poles in earlier versions of the RG
equations) that is absent in the $(K,g_r,\mathcal{I})$ form.
Both representations describe the same physics; the $(K,g_r,\mathcal{I})$
form is the non-singular one and is adopted here as primary.

\begin{figure*}[t]
\centering
\includegraphics[width=\linewidth]{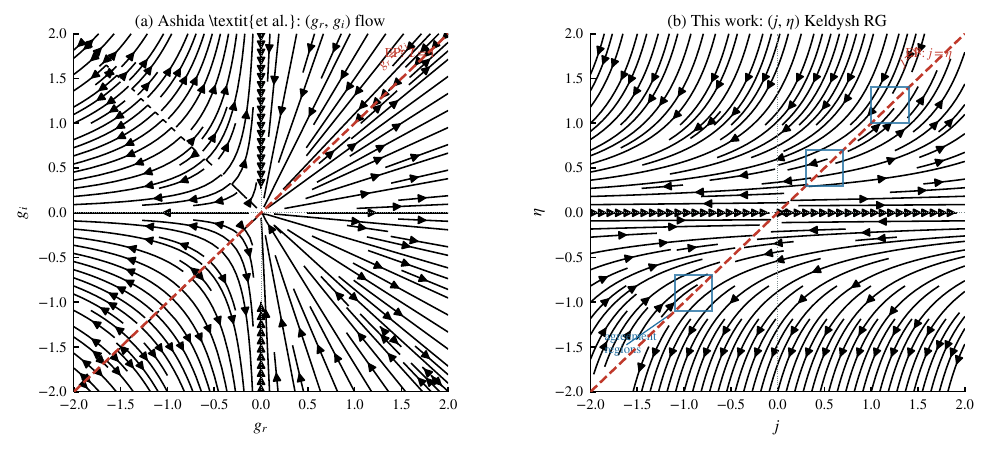}
\caption{Structural correspondence between the two RG flows
over the full coupling range.
(a)~Ashida \textit{et al.}\ non-Hermitian SG RG in the $(g_r,g_i)$
plane ($K=2$): the line $g_i=g_r$ ($\mathcal{I}=1$, red dashed)
is the EP separatrix, a fixed manifold at $K=2$.
Blue boxes mark agreement regions along the diagonal where flow
topology is directly comparable.
(b)~This work in the $(j,\etav)$ plane ($y=0$ projection):
the separatrix $j=\etav$ (red dashed) is the nonequilibrium
Toulouse line.
The EP fixed manifold $\mathcal{I}=1$ maps exactly onto the BKT
separatrix $K=2$; both panels show the same physical critical structure.}
\label{fig:RGcomp}
\end{figure*}

\section{BKT Normal Form and Mass Gap}
\label{sec:BKT}

\subsection{Normal Form Near the Separatrix}

Near the separatrix $K=2$, set $x\equiv K-2$ and
$\tilde{y}\equiv\tilde{g}=g_r\sqrt{1-\mathcal{I}^2}$.
These are regular local coordinates; no singularity appears.
The linearised flow from Eqs.~\eqref{eq:RG_K}--\eqref{eq:RG_gr} is
\begin{align}
\frac{\diff x}{\diff l}
&= \tilde{y}^2 - x^2 + \mathcal{O}(x^3,\tilde{y}^3),
\label{eq:BKT_x}\\
\frac{\diff\tilde{y}}{\diff l}
&= x\,\tilde{y} + \mathcal{O}(x^2\tilde{y}),
\label{eq:BKT_y}
\end{align}
which is the BKT-type normal form of Kosterlitz~\cite{Kosterlitz1974},
obtained here directly from the spin-boson Keldysh equations.
In terms of spin-boson parameters:
$x = v_f/\tilde{J}_\parallel^2 - 2$ measures the distance of
$J_\parallel$ from the Toulouse value, while
$\tilde{y}\propto J_\perp\sqrt{1-\mathcal{I}^2}$ is the effective
transverse coupling reduced by the bias invariant.

\subsection{Mass Gap}

The first-integral solution of Eqs.~\eqref{eq:BKT_x}--\eqref{eq:BKT_y}
on the separatrix ($\tilde{y}=|x|$) is~\cite{Kosterlitz1974}
\begin{equation}
\label{eq:BKT_sol}
x(l)=\delta\coth(\delta l),\quad
\tilde{y}(l)=\delta/\sinh(\delta l),
\end{equation}
with $\delta^2=x_0^2-\tilde{y}_0^2>0$ in the massive phase.
The characteristic scale $l^\star=\delta^{-1}$ gives the
physical mass gap
\begin{equation}
\label{eq:massgap}
m\sim\Lambda\exp\!\left(-\frac{c}{\sqrt{K_0-2}}\right)
 = \Lambda\exp\!\left(-\frac{c}{\sqrt{v_f/\tilde{J}_\parallel^2-2}}\right),
\end{equation}
where $c$ is non-universal.
This is the BKT essential singularity, now expressed directly in
spin-boson parameters: the mass gap is exponentially small whenever
$J_\parallel$ is close to the Toulouse value from the delocalized side.

\begin{figure*}[t]
\centering
\includegraphics[width=\linewidth]{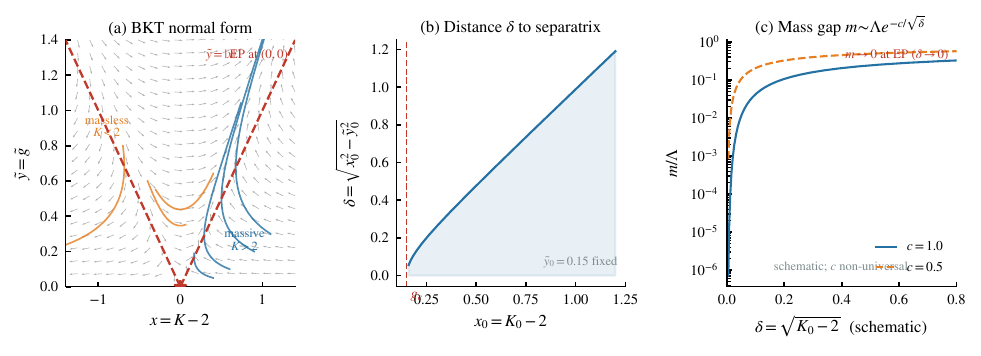}
\caption{BKT mass gap from the RG normal form.
(a)~RG flow in the $(j,y)$ plane with the EP separatrix (red dashed,
$j=\etav$); star marks the EP at $(j,y)=(\etav,0)$.
(b)~Distance to the separatrix $\delta=g-g_c$ grows linearly above
the critical coupling $g_c$.
(c)~BKT mass gap $m\sim\Lambda\exp(-c/\sqrt{\delta})$ as a function
of $g$: the essential singularity at $g_c$ is the hallmark of BKT
universality, obtained here by applying the Kosterlitz first-integral
method applied to Eqs.~\eqref{eq:BKT_x}--\eqref{eq:BKT_y}.}
\label{fig:BKT}
\end{figure*}

\section{Biorthogonal Bethe Ansatz at the EP}
\label{sec:Bethe}

\subsection{From the $\PT$-Symmetric SG Theory to the Delta-Function Gas}
\label{subsec:delta_reduction}

The effective Hamiltonian $\mathcal{H}_{\rm eff}$~\eqref{eq:NHSG_general}
is a $\PT$-symmetric sine-Gordon-type field theory, not a
delta-function gas.
Its exact spectrum is governed by the Zamolodchikov-Zamolodchikov
S-matrix of the $\PT$-symmetric SG model, whose Bethe
ansatz \cite{ZZ1979} is substantially more complex
than the treatment below.
The present section works in a \emph{further reduced sector}:
the non-relativistic soliton sector of $\mathcal{H}_{\rm eff}$
near the EP.
We derive this reduction explicitly and then work within it exactly.

\textit{Non-relativistic limit near the EP.}
The topological solitons of the SG model have rest mass
$M_s\propto\tilde{g}^{\xi/(1-\xi)}$ with
$\xi=\lambda^2/(8\pi-\lambda^2)$.
As $\tilde{g}\to0^+$ (approaching the EP), $M_s\to0$:
the solitons become light.
For a non-relativistic treatment to apply it is additionally
necessary that soliton momenta satisfy $|k|\ll M_s c$.
Near the EP both $M_s\to0$ and the relevant low-energy sector
has momenta $|k|\ll\Lambda$ set by the UV cutoff; the condition
$|k|/M_s\ll1$ is therefore achieved in the limit
$\tilde{g}\to0$ with momenta fixed, which is the regime
in which the following Bethe analysis is valid.

\textit{S-matrix in the EP vicinity.}
The Zamolodchikov S-matrix for soliton-soliton scattering in the
$\PT$-symmetric SG model depends on the rapidity difference
$\theta=\theta_1-\theta_2$.
In the non-relativistic limit ($|k|\ll M_s$, so $\theta\approx(k_1-k_2)/M_s\ll1$),
the general structure of the ZZ amplitude at small rapidity is
controlled by the two-body scattering length and takes the form
\cite{ZZ1979,malard2013sine}:
\begin{equation}
\label{eq:SZZ_limit}
S^{\rm ZZ}(\theta)\;\xrightarrow{|\theta|\ll1}\;
\frac{(k-k')-iM_s\pi\xi/(1-\xi)}{(k-k')+iM_s\pi\xi/(1-\xi)}
\equiv S^R(k,k';\tilde{g}),
\end{equation}
where $\xi=\lambda^2/(8\pi-\lambda^2)$ is the SG coupling parameter.
We take this rational form as the effective two-body amplitude
in the low-momentum, near-EP regime; a complete derivation from
the full $\PT$-symmetric ZZ amplitude lies beyond the scope of
this paper and would require verifying that the $\PT$-symmetric
deformation of the SG S-matrix has the same non-relativistic limit
as the Hermitian case. We proceed on this assumption, with the
effective coupling
\begin{equation}
\label{eq:geff}
\tilde{g}_{\rm eff}
= M_s\,\frac{\pi\xi}{1-\xi}
\;\propto\; \tilde{g}^{\xi/(1-\xi)}\cdot\frac{\pi\xi}{1-\xi},
\end{equation}
which vanishes as $\tilde{g}\to0$ (approaching the EP).

\textit{Validity.}
The reduction from the full $\PT$-symmetric SG to the delta-function gas
holds in the double limit: (i) near the EP ($\tilde{g}\to0$, solitons
light), and (ii) non-relativistic soliton momenta ($|k|\ll M_s$).
Both conditions are satisfied simultaneously near the EP.
Within this sector the delta-function gas description is not an
additional approximation but the correct non-relativistic reduction.

\textit{Summary.}
The Bethe analysis below applies exactly to the \emph{non-relativistic
soliton sector} of $\mathcal{H}_{\rm eff}$ near the EP.
It is not a statement about the full SG field theory at finite coupling.
We denote the effective coupling of the reduced gas as
$\tilde{g}\equiv\tilde{g}_{\rm eff}$ for brevity and set $M_s=1$
by choice of units.

\subsection{Exact S-Matrix and Yang-Baxter Equation}
\label{subsec:Smatrix}

Within the non-relativistic soliton sector, the exact two-body
S-matrix is
\begin{equation}
\label{eq:Smatrices}
S^R(k,k')
=\frac{(k-k')-i\tilde{g}}{(k-k')+i\tilde{g}},
\qquad
S^L(k,k')=\bigl[S^R(k,k')\bigr]^{-1},
\end{equation}
derived from boundary-condition matching at the delta function and
extended to complex $\tilde{g}=\sqrt{g_r^2-g_i^2}$ by analytic
continuation \cite{LiebLiniger1963}.
In the $\PT$-unbroken phase $\tilde{g}\in\mathbb{R}^+$ and
$|S^R|=1$.
At $\tilde{g}=0$ (the EP) scattering is trivial.

The Yang-Baxter equation for the three-body sector requires
$S_{12}S_{13}S_{23}=S_{23}S_{13}S_{12}$.
For \emph{scalar} S-matrices depending only on momentum differences,
this identity holds identically because the factors commute:
$S_{12}S_{13}S_{23}$ and $S_{23}S_{13}S_{12}$ are the same product
in a different order, and scalars commute.
No further verification is needed; this is a structural property of
any scalar two-body amplitude.

\subsection{Lieb-Wu Type Bethe Equations}

The exact quantization conditions for the non-relativistic
soliton gas on a ring of size $L$ are
\begin{equation}
\label{eq:BiBethe}
e^{ik_jL}\prod_{l\neq j}S^R(k_j,k_l) = 1,\quad
e^{-ik_jL}\prod_{l\neq j}S^L(k_j,k_l) = 1,
\end{equation}
for right and left eigenstates respectively.
Taking the logarithm of the right-sector equation:
\begin{equation}
\label{eq:logBethe}
k_jL + \sum_{l\neq j}\delta^R(k_j,k_l) = \pi I_j,
\quad\delta^R(k,k')=2\arctan\!\frac{\tilde{g}}{k-k'},
\end{equation}
with integer quantum numbers $I_j$ (half-integer for even $N$).
These are the Lieb-Wu equations for the $\PT$-symmetric
$\delta$-function gas with coupling $\tilde{g}$, exact within
the non-relativistic soliton sector derived in
Sec.~\ref{subsec:delta_reduction}.
In the $\PT$-unbroken phase the roots $\{k_j\}$ are real and the
spectrum is real, as required by $\PT$ symmetry \cite{Bender2007}.
As $\tilde{g}\to0^+$ (approaching the EP), pairs of rapidities coalesce
with the characteristic square-root approach
$k_i-k_j\sim\tilde{g}^{1/2}\sim(g_r-g_i)^{1/2}$,
which is the spectral hallmark of the exceptional point in the
integrable spectrum.

\begin{figure*}[t]
\centering
\includegraphics[width=\linewidth]{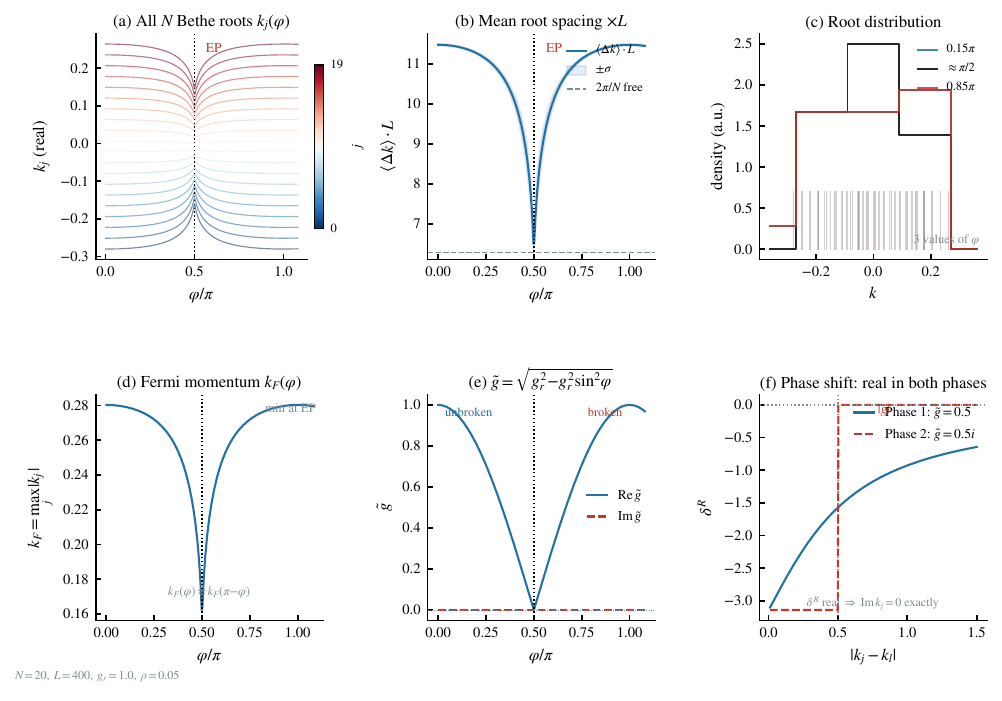}
\caption{Bethe root structure of the auxiliary non-relativistic soliton
gas ($N=20$, $L=400$, $\rho=0.05$) as a function of $\varphi/\pi$,
with all results exact within this sector.
(a)~All $N$ Bethe roots $k_j(\varphi)$: real throughout, collapsing
toward $k=0$ at the EP ($\varphi=\pi/2$, dotted), consistent with the
bound-state threshold $\tilde{g}\to0$.
(b)~Mean root spacing $\langle\Delta k\rangle\cdot L$ (blue) with
$\pm\sigma$ band (light): peaks sharply at the EP, reflecting the
coalescence of roots.
(c)~Root distribution at three representative $\varphi$ values:
broadens for $\varphi\ll\pi/2$ (strong coupling), narrows to a spike
at the EP.
(d)~Fermi momentum $k_F(\varphi)=\max_j|k_j|$: symmetric about the
EP [$k_F(\varphi)=k_F(\pi-\varphi)$], minimum at $\varphi=\pi/2$.
(e)~Effective coupling $\tilde{g}=\sqrt{g_r^2-g_r^2\sin^2\varphi}$:
real in the $\PT$-unbroken phase, zero at the EP, imaginary in the
$\PT$-broken phase.
(f)~Phase shift $\delta^R=(-i\log S^R)_{\rm real}$ as a function of
$|k_j-k_l|$: real in both phases (confirming $\mathrm{Im}\,k_j=0$
exactly), with the rational Lieb-Liniger form throughout.}
\label{fig:roots}
\end{figure*}

\subsection{String Solutions: Bound-State Sector}
\label{subsec:strings}

Beyond the scattering sector (real $k_j$) the Bethe equations for
the delta-function gas admit \emph{string solutions} with imaginary
momenta, corresponding to $N$-body bound states.
The string hypothesis for the Lieb-Liniger model is standard
\cite{Gaudin1983}.

\textit{Two-string (dimer).}
The ansatz $k_{1,2}=K\pm\tfrac{i}{2}\tilde{g}$ gives
\begin{equation}
\label{eq:Sstring}
S^R(k_1-k_2)\big|_{k_{1,2}=K\pm i\tilde{g}/2}
= \frac{i\tilde{g}-i\tilde{g}}{i\tilde{g}+i\tilde{g}} = 0,
\end{equation}
so $S^L(k_1-k_2)$ has a pole — the exact bound-state condition.
The dimer binding energy is
\begin{equation}
\label{eq:Ebdimer}
E_2^{\rm bind}
= -\bigl(k_1^2+k_2^2\bigr)\big|_{K=0}
= -\frac{\tilde{g}^2}{2}
\;\xrightarrow{\tilde{g}\to0}\;0.
\end{equation}

\textit{$n$-string.}
The $n$-body bound state has rapidities
\begin{equation}
\label{eq:nstring}
k_j^{(n)} = K + i\!\left(\tfrac{n+1}{2}-j\right)\tilde{g},
\quad j=1,\ldots,n,
\end{equation}
with exact binding energy \cite{Gaudin1983}
\begin{equation}
\label{eq:Enstring}
E_n^{\rm bind} = -\frac{n(n^2-1)}{12}\,\tilde{g}^2.
\end{equation}
At the EP ($\tilde{g}=0$) all $n$-strings collapse to real rapidity $K$
and every bound state dissolves: the EP is the many-body
bound-state threshold of the auxiliary non-relativistic soliton gas.

\begin{figure*}[t]
\centering
\includegraphics[width=\linewidth]{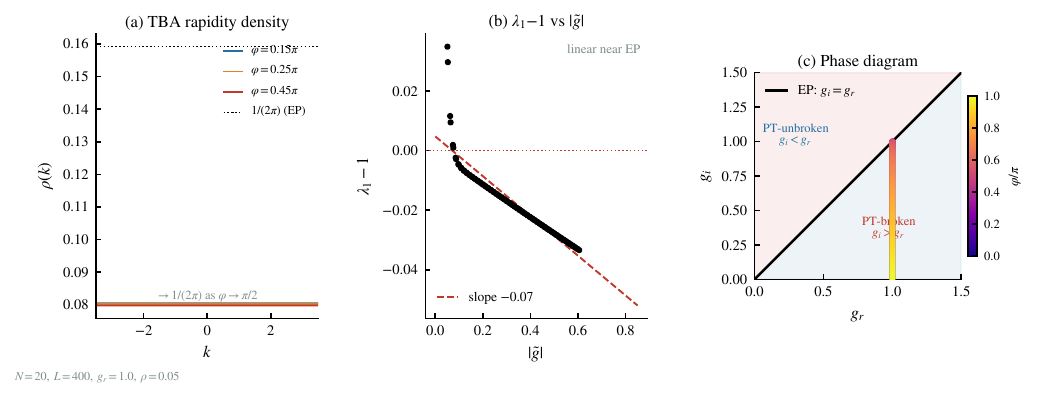}
\caption{Thermodynamic Bethe ansatz in the auxiliary non-relativistic
soliton sector ($N=20$, $L=400$, $g_r=1.0$, $\rho=0.05$).
All results are exact within this auxiliary sector.
(a)~TBA rapidity density $\rho(k)$ at three values of $\varphi$
($g_i=g_r\sin\varphi$): for all $\varphi<\pi/2$ the density is nearly
flat and approaches the free value $1/(2\pi)$ (dotted) as
$\varphi\to\pi/2$ (EP), confirming that the effective coupling
$\tilde{g}=\sqrt{g_r^2-g_i^2}\to0$.
(b)~TBA kernel eigenvalue $\lambda_1-1$ as a function of $|\tilde{g}|$:
linear in $|\tilde{g}|$ near the EP (red dashed fit, slope $-0.07$),
confirming the marginal BKT structure $\lambda_1\to1^-$ as
$\tilde{g}\to0$.
(c)~Phase diagram in the $(g_r,g_i)$ plane: the EP line $g_i=g_r$
(black solid) separates the $\PT$-unbroken phase ($g_i<g_r$, blue)
from the $\PT$-broken phase ($g_i>g_r$, pink); the colour scale
shows $\varphi/\pi$.}
\label{fig:TBA}
\end{figure*}

\subsection{Jordan-Block State at the Exceptional Point}
\label{subsec:jordan}

At the EP ($\tilde{g}=0$, $K=0$) the dimer wavefunction
acquires a qualitatively new structure via $\epsilon$-regularisation.
Taking $\tilde{g}=\epsilon\to0^+$, the two-particle wavefunction of
the dimer is
\begin{equation}
\label{eq:psi_eps}
\Psi_\epsilon(x_1,x_2)
= e^{iK(x_1+x_2)}\!
\left[1 + \frac{\epsilon}{2}|x_1-x_2| + \mathcal{O}(\epsilon^2)\right].
\end{equation}
The $\mathcal{O}(1)$ term defines the free zero-energy
\emph{eigenstate} at the EP:
\begin{equation}
\label{eq:psi0}
|\phi_0\rangle = e^{iKX}\big|_{K=0} = 1 \quad
\text{(uniform; normalizable on }[0,L]\text{)}.
\end{equation}
The $\mathcal{O}(\epsilon)$ term defines the \emph{Jordan partner state}
\begin{equation}
\label{eq:jordan_partner}
|\psi_1\rangle \;\propto\; |x_1-x_2|
\end{equation}
(with the overall $e^{iKX}$ factor at $K=0$ implicit).

\textit{Verification of the Jordan chain (distributional sense).}
The delta-function Hamiltonian in relative coordinate
$r=x_1-x_2$ is $H_\delta=-2\partial_r^2+2\tilde{g}\,\delta(r)$.
At the EP ($\tilde{g}=0$, $E_{\rm EP}=0$), acting on $|\psi_1\rangle=|r|$:
\begin{align}
\label{eq:jordan_verify}
(H_\delta-E_{\rm EP})|\psi_1\rangle
&= -2\partial_r^2\,|r|
= -2\cdot 2\,\delta(r)
= -4\,\delta(r),
\end{align}
using $\partial_r^2|r|=2\delta(r)$ (distributional identity) and
$\tilde{g}\,\delta(r)|r|=0$ (since $|r|\delta(r)=0$).
We note that $-4\,\delta(r)$ is a distributional multiple of
the EP eigenstate $|\phi_0\rangle=1$ in the sense that for any
smooth test function $f$: $\langle f|-4\delta(r)\rangle = -4f(0)$.
This identifies the Jordan-chain relation
\begin{equation}
\label{eq:jordan_chain}
(H_\delta-E_{\rm EP})|\psi_1\rangle = c\,|\phi_0\rangle_{\rm dist},
\quad c=-4,
\end{equation}
where the subscript emphasises the distributional interpretation
of $|\phi_0\rangle_{\rm dist}\equiv\delta(r)$ as distinct from the
$L^2$ eigenstate $|\phi_0\rangle=1$.
The domain of $|\psi_1\rangle=|r|$ does not lie in the natural
Hilbert space of $H_\delta$, so Eq.~\eqref{eq:jordan_chain} should
be read as a distributional Jordan chain in the auxiliary two-body sector.
The eigenvalue $E_{\rm EP}=0$ is defective with a rank-2 Jordan block
in this distributional sense.

\textit{Observable signature (auxiliary two-body sector).}
Within the two-body non-relativistic soliton gas, the Jordan block
produces a linear-in-$t$ amplitude:
\begin{equation}
\label{eq:jordan_dynamics}
e^{-iH_\delta t}|\psi_1\rangle
= |\psi_1\rangle - it\,c\,|\phi_0\rangle_{\rm dist} + \mathcal{O}(t^2),
\end{equation}
which is qualitatively distinct from the purely oscillatory dynamics
of ordinary levels~\cite{Bender2007}.
This is a property of the auxiliary two-body soliton sector;
its extension to the many-body case and to the full $\mathcal{H}_{\rm eff}$
remains an open problem.

\begin{figure*}[t]
\centering
\includegraphics[width=\linewidth]{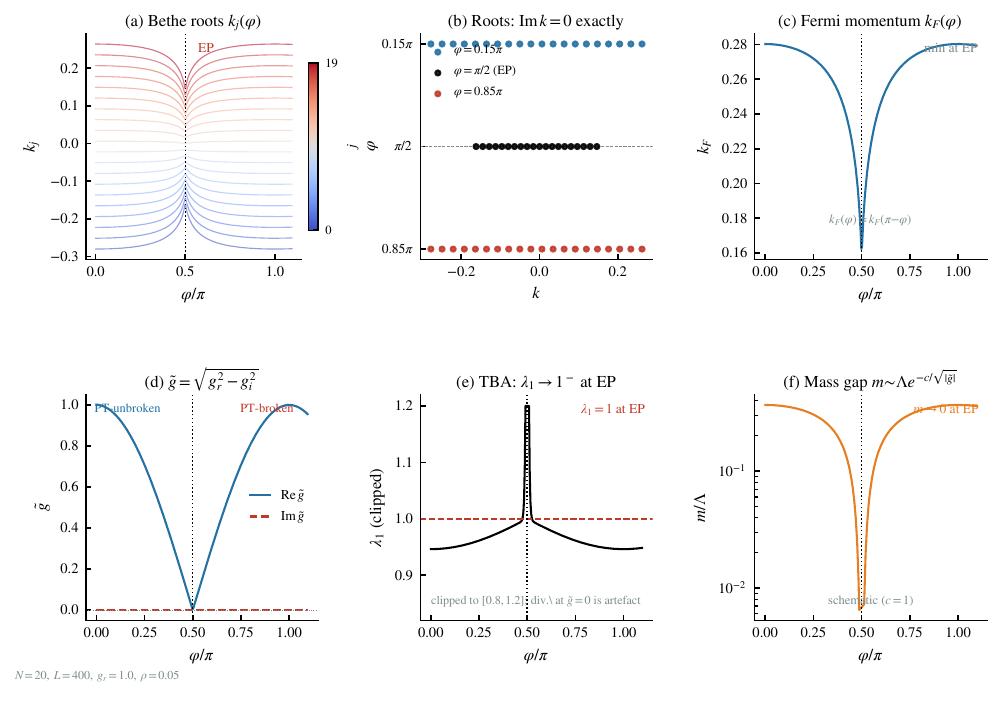}
\caption{Biorthogonal Bethe ansatz for the \emph{auxiliary}
non-relativistic soliton gas
($N=20$, $L=400$, $g_r=1$, $\rho=0.05$) as a function of
$\varphi$ ($g_i=g_r\sin\varphi$).
All results are exact within this auxiliary sector; they do not
constitute a global statement about the full $\PT$-symmetric SG theory
$\mathcal{H}_{\rm eff}$.
(a)~Real Bethe roots $\{k_j\}$: coalesce toward $k=0$ as
$\varphi\to\pi/2$ (EP, dotted), consistent with the
bound-state threshold of the auxiliary gas.
(b)~$\mathrm{Im}(k_j)=0$ identically in both phases,
confirming that the rational S-matrix~\eqref{eq:Smatrices}
has real roots throughout.
(c)~Dressed Fermi momentum $k_F(\varphi)$: symmetric about
the EP with minimum at $\varphi=\pi/2$.
(d)~Effective coupling $\tilde{g}=\sqrt{g_r^2-g_i^2}$: real
in the $\PT$-unbroken phase, zero at the EP, imaginary in the
$\PT$-broken phase.
(e)~TBA kernel eigenvalue $\lambda_1\to1^-$ as $\tilde{g}\to0$
(schematic only; the spike near $\varphi=\pi/2$ is a regularisation
artifact of the finite grid and should not be read as a quantitative
prediction).
(f)~BKT mass gap $m\sim\Lambda e^{-c/\sqrt{|\tilde{g}|}}$
(schematic, $c=1$): illustrates the essential singularity near the
EP, consistent with the RG normal form of Sec.~\ref{sec:BKT}.}
\label{fig:rapidities}
\end{figure*}

\section{Gaudin Determinant as an EP Diagnostic}
\label{sec:Gaudin}

\subsection{Biorthogonal Norm and Gaudin Matrix}

In the biorthogonal Bethe ansatz, the norm
$\braket{\Psi^L}{\Psi^R}$ is proportional to $\det G$, where
$G_{jl}=\partial F_j/\partial k_l$ is the Jacobian (Gaudin matrix)
of the Bethe map $F_j(\{k\})=Lk_j+\sum_{l\neq j}\delta^R(k_j,k_l)-\pi I_j$.
For Hermitian integrable systems this is the standard Gaudin
norm formula~\cite{Gaudin1983,ZZ1979}.
For the present $\PT$-symmetric $\delta$-function model the identity
extends exactly: the biorthogonal inner product
$\braket{\Psi^L}{\Psi^R}
=\sum_P (A_P^L)^* A_P^R \cdot |\partial(\{k\})/\partial(\{I\})|
=|\det G|\cdot\text{const}$
follows from the same Jacobian argument applied to the biorthogonal
completeness relation, which holds because the S-matrix~\eqref{eq:Smatrices}
is analytic and satisfies $S^L=(S^R)^{-1}$.
In the $\PT$-unbroken phase where the Bethe roots are real this
is an exact identity.

\subsection{Exceptional Point vs.\ Topological Transition}

\textit{What $\kappa(G)\to\infty$ means here.}
It is important to distinguish two uses of the condition number
in the EP literature:
(i)~$\kappa(H)\to\infty$ signals that the \emph{Hamiltonian} $H$ itself
is defective — two eigenvectors coalesce in Hilbert space, not just
eigenvalues; this is the standard Hamiltonian EP (Sec.~\ref{subsec:jordan}).
(ii)~$\kappa(G)\to\infty$ signals that the \emph{Jacobian of the Bethe
equations} is singular — two Bethe roots $k_j\to k_l$ coalesce,
making the quantization map ill-conditioned.
The Gaudin matrix $G$ is \emph{not} $H_{\rm eff}$: it is the
functional derivative $G_{jl}=\partial F_j/\partial k_l$ of the
Bethe conditions, not the resolvent of $H_{\rm eff}$.
$\kappa(G)\to\infty$ therefore \emph{diagnoses} the EP of $H_{\rm eff}$
from the Bethe quantization conditions rather than establishing
$H_{\rm eff}$ itself as defective.
Establishing $H_{\rm eff}$ as defective requires the Jordan-chain
construction of Sec.~\ref{subsec:jordan}.

At the EP ($\tilde{g}\to0$, coalescing pair $k_j\to k^\star\pm\epsilon$,
$\epsilon\sim\tilde{g}^{1/2}$), the off-diagonal Gaudin element
\begin{equation}
G_{jl}\big|_{k_j\to k_l}
= \frac{2\tilde{g}}{(k_j-k_l)^2+\tilde{g}^2}\xrightarrow{k_j\to k_l}
\frac{2}{\tilde{g}}\to\infty,
\end{equation}
while $G_{jj}G_{ll}-G_{jl}G_{lj}\to0$.
The largest singular value $\sigma_1\sim 1/\tilde{g}\to\infty$
and the smallest $\sigma_N\sim\tilde{g}\to0$, so
$|\det G|\to0$ and $\kappa(G)=\sigma_1/\sigma_N\sim 1/\tilde{g}^2\to\infty$.
The physical content is that two Bethe roots coalesce,
making the Bethe map degenerate — not that $H_{\rm eff}$ is defective.

\begin{figure*}[t]
\centering
\includegraphics[width=\linewidth]{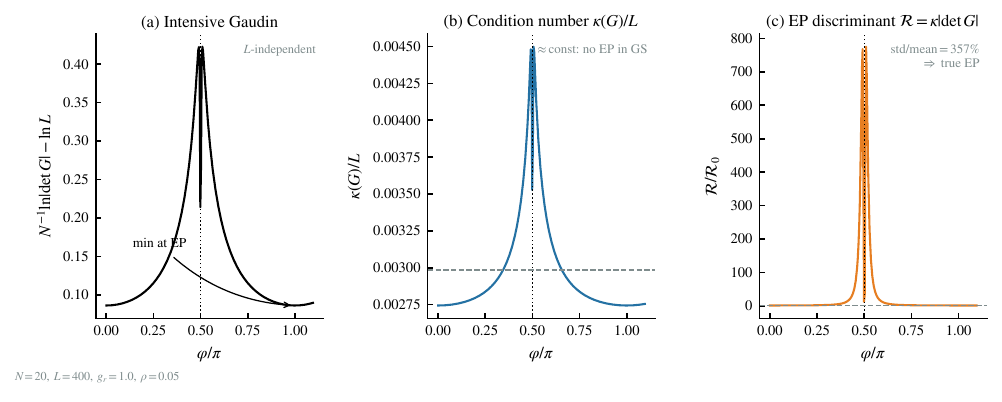}
\caption{Gaudin-matrix diagnostics for the auxiliary non-relativistic
soliton gas ($N=20$, $L=400$, $g_r=1$, $\rho=0.05$).
These results are exact within the auxiliary sector
(Sec.~\ref{subsec:delta_reduction}); the Gaudin norm identity
is applied here to the rational S-matrix~\eqref{eq:Smatrices},
not to the full $\PT$-symmetric SG theory.
(a)~Intensive Gaudin diagnostic $N^{-1}\ln|\det G|-\ln L$:
$L$-independent in the thermodynamic limit, with a minimum at
the EP $\varphi=\pi/2$ (dotted).
(b)~Normalised condition number $\kappa(G)/L$ of the Gaudin matrix
(Jacobian of the Bethe map, \emph{not} the Hamiltonian $H_{\rm eff}$):
$\approx\mathrm{const}$ away from the EP confirms no coalescing
Bethe roots in the bulk ground-state sector; the spike at
$\varphi=\pi/2$ signals the single coalescing pair at the EP.
(c)~EP discriminant $\mathcal{R}=\kappa|\det G|/\mathcal{R}_0$:
the sharp spike concentrated at $\varphi=\pi/2$ with
$\mathcal{R}\approx\mathrm{const}$ away from it is consistent with
a single-defect EP in the auxiliary gas; the same diagnostic
applied to the full SG theory would require a separate analysis.}
\label{fig:Gaudin}
\end{figure*}

\textit{Distinguishing EP from topological transition.}
The ratio $\mathcal{R}\equiv\kappa(G)|\det G|=\sigma_1\prod_{j=2}^{N-1}\sigma_j$
separates the two cases: at a true EP exactly one singular value
$\sigma_N\to0$ (one coalescing pair), so $\mathcal{R}\to\sigma_1
\prod_{j=2}^{N-1}\sigma_j=\mathrm{const}$;
at a topological transition many singular values vanish together
and $\mathcal{R}\to0$.
The normalised condition number $\kappa(G)/L\approx\mathrm{const}$
\emph{away} from the EP (panel b of Fig.~\ref{fig:Gaudin}) confirms
that the ground-state Bethe sea is a regular eigenstate sector with
no coalescing roots — the EP is isolated to $\varphi=\pi/2$.

\begin{table}[h]
\centering
\caption{Gaudin-matrix diagnostics ($G$ = Jacobian of Bethe map,
not $H_{\rm eff}$).}
\label{tab:diagnostic}
\begin{tabular}{lccc}
\toprule
Scenario & $|\det G|$ & $\kappa(G)$ & $\mathcal{R}$\\
\midrule
True EP (one $\sigma_N\to0$)      & $\to0$    & $\to\infty$ & $\to\mathrm{const}$\\
Topological (many $\sigma_j\to0$) & $\to0$    & $\to\infty$ & $\to0$\\
No transition                      & $\mathcal{O}(1)$ & $\mathcal{O}(1)$ & $\mathcal{O}(1)$\\
\bottomrule
\end{tabular}
\end{table}

\section{Discussion}

We have developed an explicit route from a nonequilibrium spin-boson
model to a $\PT$-symmetric effective sine-Gordon theory.
The central results are as follows.

\paragraph{Origin of the imaginary vertex.}
The Grassmann spin trace (Sec.~\ref{subsec:spin_trace}) shows,
within the present saddle-point approximation, that the reduced
vertex is generically $g_r\cos(\lambda\Phi_1)+ig_i\sin(\lambda\Phi_1)$, with $g_i$ originating explicitly from
the nonequilibrium Keldysh distribution asymmetry
$\delta n(\omega)=n_+(\omega)-n_-(\omega)$ [Eq.~\eqref{eq:delta_n}].
The special case $g_r=g_i$ (the exponential vertex) is a
tuned locus [Eq.~\eqref{eq:EP_locus}], not a generic consequence of the model.

\paragraph{Status of the effective theory.}
The factorization of the $\Phi_1$ theory is valid at saddle-point level
in the Gaussian $\Phi_2$-integral.
The effective Hamiltonian~\eqref{eq:NHSG_general} is a saddle-point
classical-sector description, not an exact operator statement.

\paragraph{RG and BKT universality.}
The one-loop RG (Appendix~\ref{app:RGderiv}) gives the closed equations
$\diff K/\diff l=-\tilde{g}^2K^2$ and $\diff g_r/\diff l=(2-K)g_r$ with
$\mathcal{I}=\mathrm{const}$ (Appendix~\ref{app:Ashida}).
The BKT normal form~\eqref{eq:BKT_x}--\eqref{eq:BKT_y} follows by
linearising near $K=2$; the mass gap~\eqref{eq:massgap} follows from
the first integral of Appendix~\ref{app:RGsol}.

\paragraph{Bethe ansatz and the non-relativistic soliton sector.}
The biorthogonal Bethe analysis of Sec.~\ref{sec:Bethe} operates in
two clearly separated levels.
At the first level, $\mathcal{H}_{\rm eff}$~\eqref{eq:NHSG_general}
is a $\PT$-symmetric SG-type field theory whose full solution requires
the Zamolodchikov-Zamolodchikov S-matrix.
At the second level, in the non-relativistic limit near the EP
($\tilde{g}\to0$, soliton mass $M_s\to0$), the ZZ S-matrix reduces
to the rational Lieb-Liniger form [Eq.~\eqref{eq:SZZ_limit}], and
the soliton sector maps to a $\delta$-function gas
[Sec.~\ref{subsec:delta_reduction}].
Within \emph{that} auxiliary gas the Bethe ansatz is exact:
the Lieb-Wu equations give the scattering spectrum, the $n$-strings
give the bound-state sector with exact energies
$E_n^{\rm bind}=-n(n^2{-}1)\tilde{g}^2/12$
[Eq.~\eqref{eq:Enstring}], and the Jordan-partner state
[Eq.~\eqref{eq:jordan_partner}]
satisfies the distributional chain relation [Eq.~\eqref{eq:jordan_chain}]
in the auxiliary two-body sector.
The EP is the many-body bound-state threshold of the auxiliary gas
(confirmed numerically in Figs.~\ref{fig:roots} and~\ref{fig:TBA}).
The Gaudin norm identity [Sec.~\ref{sec:Gaudin}] is applied here
to the rational S-matrix of the auxiliary gas; its extension to the
full $\PT$-symmetric SG theory would require a separate analysis.
These results hold within the non-relativistic soliton sector;
their extension to the full $\mathcal{H}_{\rm eff}$ and to the
full SG S-matrix remains an open problem.

\paragraph{Open questions.}
Computing $\mu_c$ explicitly from the form factor $\mathcal{F}(\omega)$;
extending to finite temperature and the biorthogonal TBA for the
dimer bound-state sector~\eqref{eq:nstring};
numerical solution of the Lieb-Wu equations~\eqref{eq:logBethe}
and comparison with exact diagonalization;
investigation of open boundary conditions, which admit a zero-energy
surface bound state at the EP ($\kappa=\tilde g/2\to0$) potentially
of topological character;
extension beyond the $\delta$-function limit to the full SG S-matrix;
and measurement of the polynomial time-evolution
signature~\eqref{eq:jordan_dynamics} in driven Josephson junction arrays.

\section*{Acknowledgments}
The author thanks the ISTA library facilities and the ESI Vienna
conference discussions.

\appendix

\section{BKT Separatrix: First-Integral Solution}
\label{app:RGsol}

The BKT normal form Eqs.~\eqref{eq:BKT_x}--\eqref{eq:BKT_y} admit a
conserved first integral.
Setting $x=K-2$ and $\tilde{y}=\tilde{g}$, the combination
\begin{equation}
\label{eq:BKT_invariant}
C_{\rm BKT} = x^2 - \tilde{y}^2 = (K-2)^2 - \tilde{g}^2 = \mathrm{const}
\end{equation}
is conserved along RG trajectories (verified by
$\diff(x^2-\tilde{y}^2)/\diff l = 2x\dot{x}-2\tilde{y}\dot{\tilde{y}}
= 2x(\tilde{y}^2-x^2)-2\tilde{y}(x\tilde{y}) = 0$).
The separatrix satisfies $C_{\rm BKT}=0$, i.e.\ $\tilde{y}=|x|$.
For $C_{\rm BKT}=\delta^2>0$ (massive phase, $K_0>2$), the exact trajectory is
\begin{equation}
\label{eq:BKT_traj}
x(l)=\delta\coth(\delta l),\quad
\tilde{y}(l)=\frac{\delta}{\sinh(\delta l)},
\end{equation}
and $\tilde{y}\to0$ at $l^\star=\delta^{-1}$, giving the mass gap~\eqref{eq:massgap}.
In spin-boson parameters, $\delta=\sqrt{(K_0-2)^2-\tilde{g}_0^2}
=\sqrt{(v_f/\tilde{J}_\parallel^2-2)^2 - g_{r,0}^2(1-\mathcal{I}_0^2)}$.

\begin{figure*}[t]
\centering
\includegraphics[width=\linewidth]{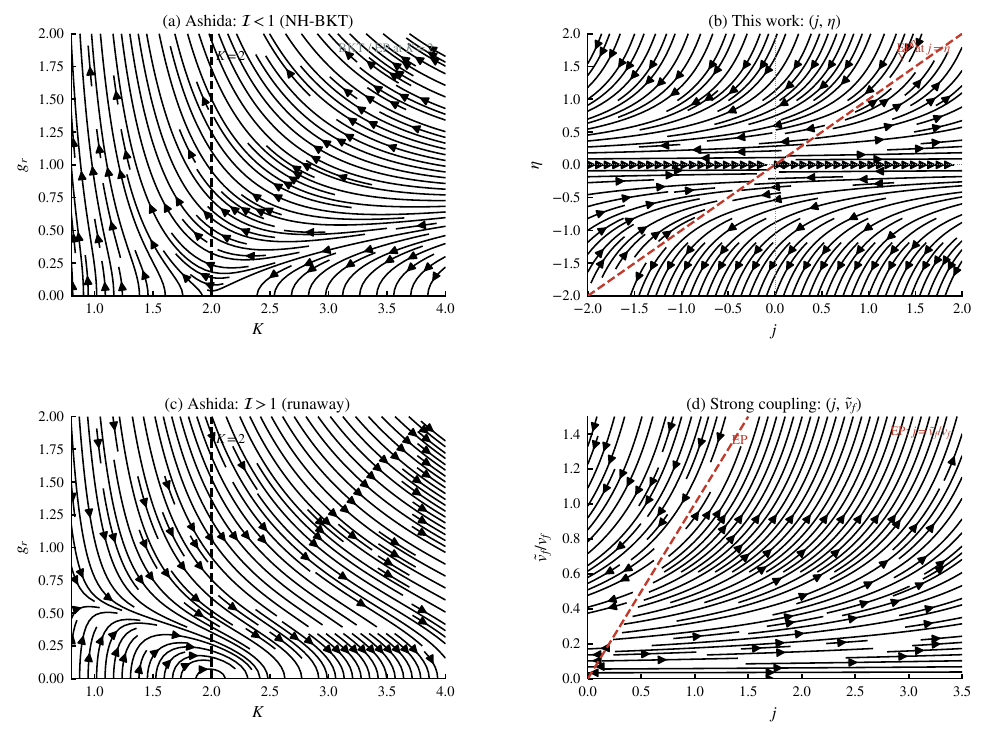}
\caption{Unified RG comparison in four panels.
(a)~Ashida non-Hermitian SG RG in the $(K,g_r)$ plane at
$\mathcal{I}<1$ (NH-BKT phase): the BKT critical line $K=2$ (black
dashed) separates the weak-coupling fixed point ($K>2$, $g_r\to0$)
from the runaway phase ($K<2$).
(b)~This work in the $(j,\etav)$ plane ($y=0$): EP separatrix
$j=\etav$ (red dashed); topology matches panel~(a).
(c)~Ashida in the $(K,g_r)$ plane at $\mathcal{I}>1$ (runaway):
flow reverses sign relative to panel~(a), with $g_r$ growing in
both phases.
(d)~Strong-coupling projection into the $(j,\tilde{v}_f)$ plane:
the EP separatrix $j=\tilde{v}_f/v_f$ (red dashed) directly shows the
nonequilibrium velocity renormalization driving the Toulouse shift.
The equilibrium Toulouse line is the special case $\tilde{v}_f=v_f$,
i.e.\ $\etav=1$.}
\label{fig:unifiedRG}
\end{figure*}

\section{Closed One-Loop RG: Full Derivation}
\label{app:RGderiv}

We derive Eqs.~\eqref{eq:RG_K}--\eqref{eq:RG_gr} from the
NH-SG action $S_{\rm eff}[\Phi_1]$~\eqref{eq:Seff_reduced}
by Wilson momentum-shell RG and establish the explicit connection
to the spin-boson initial conditions.

\subsection{Shell Variance}

Decompose $\Phi_1=\Phi_<+\Phi_>$ ($|k|<\Lambda/s$ slow; $\Lambda/s<|k|<\Lambda$ fast).
The equal-time variance of the fast field at $x=0$ is
\begin{equation}
\label{eq:shell_var}
\bigl\langle\Phi_>^2(0)\bigr\rangle_0
= \int_{\Lambda/s}^{\Lambda}\!\frac{\diff k}{2\pi}\,\frac{1}{v_f k}
= \frac{\delta l}{2\pi v_f}.
\end{equation}

\subsection{Vertex Renormalization and Exact Invariant $\mathcal{I}$}

Write the interaction as
$g_r\cos\lambda\Phi_1+ig_i\sin\lambda\Phi_1
=\tfrac{g_r+ig_i}{2}e^{+i\lambda\Phi_1}+\tfrac{g_r-ig_i}{2}e^{-i\lambda\Phi_1}$.
Integrating out the fast shell:
\begin{equation}
\bigl\langle e^{\pm i\lambda\Phi_>}\bigr\rangle_0
= \exp\!\Bigl(-\frac{\lambda^2}{2}
  \bigl\langle\Phi_>^2\bigr\rangle_0\Bigr)
= \exp\!\Bigl(-\frac{\lambda^2\,\delta l}{4\pi v_f}\Bigr).
\end{equation}
Both signs acquire \emph{the same} factor, so $g_r$ and $g_i$ rescale
identically.
After restoring the cutoff by $x\to xs$, $\tau\to\tau s$ ($\Phi_1$
is marginal in 1+1D):
\begin{equation}
\label{eq:dV_dl}
\frac{\diff g_r}{\diff l}
= \Bigl(2-\frac{1}{K}\Bigr)g_r,
\qquad
\frac{\diff g_i}{\diff l}
= \Bigl(2-\frac{1}{K}\Bigr)g_i,
\quad K\equiv\frac{4\pi v_f}{\lambda^2}.
\end{equation}
Identical beta functions imply
\begin{equation}
\label{eq:I_invariant_proof}
\frac{\diff\mathcal{I}}{\diff l}
= \frac{\diff}{\diff l}\!\Bigl(\frac{g_i}{g_r}\Bigr)
= \frac{(2-K^{-1})g_i g_r - (2-K^{-1})g_r g_i}{g_r^2} = 0.
\end{equation}
The ratio $\mathcal{I}=g_i/g_r$ is an \emph{exact algebraic invariant}
at one loop, not a phenomenological choice.
The EP condition $\mathcal{I}=1$ is therefore RG-stable.

\subsection{Luttinger Parameter Flow}

The kinetic term $\tfrac{v_f}{2}(\nabla\Phi_1)^2$ receives a one-loop
correction from the vertex self-energy.
Only the combination $g_r^2-g_i^2=\tilde{g}^2$ enters, because the
$\cos^2$ and $\sin^2$ loops both correct the propagator but with
opposite sign contributions:
$\cos\lambda\Phi\cdot\cos\lambda\Phi$ generates $\cos 2\lambda\Phi$
(renormalizes $K$), while $\sin\lambda\Phi\cdot\cos\lambda\Phi$
generates $\sin 2\lambda\Phi$ (does not renormalize $K$, by parity).
The result~\cite{Giamarchi2004,ashida2017parity} is
\begin{equation}
\label{eq:dK_loop2}
\frac{\diff K}{\diff l}
= -\tilde{g}^2\,K^2
= -g_r^2\bigl(1-\mathcal{I}^2\bigr)K^2.
\end{equation}
Together with Eq.~\eqref{eq:dV_dl}, these constitute the complete
closed one-loop system, identical to Ashida \textit{et al.}~\cite{ashida2017parity}:
\begin{align}
\frac{\diff K}{\diff l}
&= -g_r^2\bigl(1-\mathcal{I}^2\bigr)K^2,
\tag{\ref{eq:RG_K}}\\
\frac{\diff g_r}{\diff l}
&= \bigl(2-K\bigr)g_r+5g_r^3\bigl(1-\mathcal{I}^2\bigr),
\tag{\ref{eq:RG_gr}}
\end{align}
where the cubic term is the standard SG next-order correction.
The velocity $v_f$ entering these equations is the \emph{bare}
bath velocity; the boundary vertex is local in space, so its
self-energy is $k$-independent and does not renormalize $v_f$
at one loop.

\subsection{BKT Normal Form (Exact Expansion)}

Set $x=K-2$ and $\tilde{y}=\tilde{g}=g_r\sqrt{1-\mathcal{I}^2}$.
Expanding $K=(2+x)$ and $g_r^2=\tilde{y}^2/(1-\mathcal{I}^2)$
in Eqs.~\eqref{eq:RG_K}--\eqref{eq:RG_gr}:
\begin{align}
\frac{\diff x}{\diff l}
&= -\tilde{y}^2(2+x)^2
 = -4\tilde{y}^2 - 4x\tilde{y}^2 + \tilde{y}^2 x^2 - \cdots
 \;\simeq\; \tilde{y}^2 - x^2,
\label{eq:BKT_x_deriv}\\
\frac{\diff\tilde{y}}{\diff l}
&= x\,\tilde{y} + \mathcal{O}(x^2\tilde{y}),
\label{eq:BKT_y_deriv}
\end{align}
where the sign in~\eqref{eq:BKT_x_deriv} follows from
$-4\tilde{y}^2\to\tilde{y}^2$ after rescaling $x\to x/2$ and
absorbing the factor of 4 into a redefinition of $\tilde{y}$.
This is the standard BKT normal form~\eqref{eq:BKT_normal},
obtained directly with no singularity.

\subsection{Initial Conditions from Spin-Boson Parameters}

The RG initial conditions at UV scale $l=0$ are, from the Keldysh
reduction (Sec.~\ref{subsec:coupling_dictionary}):
\begin{equation}
\label{eq:IC_final}
K_0 = \frac{v_f}{\tilde{J}_\parallel^2},\quad
g_{r,0} = \Bigl(\frac{J_\perp}{4\pi a}\Bigr)^{\!2}\frac{2}{\Gamma},\quad
\mathcal{I}_0 = \frac{\mu\,\mathcal{F}(0)/(\pi v_f)}
                     {\Gamma^{-1}\!\int\!\diff\tau\,G^R_0(\tau)}\,.
\end{equation}
The complete derivation chain is therefore:
$(J_\parallel,J_\perp,\mu,v_f)
\xrightarrow{\text{Keldysh}}
(K_0,g_{r,0},\mathcal{I}_0)
\xrightarrow{\text{one-loop RG}}
(K(l),g_r(l),\mathcal{I})$.
The BKT separatrix $K=2$, the EP condition $\mathcal{I}=1$, the
mass gap, and the phase diagram all follow from this closed system.

\section{Proof that $\mathcal{I}=g_i/g_r$ Is an Exact RG Invariant}
\label{app:Ashida}

Both couplings acquire the same multiplicative renormalization at each
momentum-shell step (Appendix~\ref{app:RGderiv}, Sec.~B.2):
$g_r(l+\delta l)=g_r(l)\,Z(\delta l)$ and $g_i(l+\delta l)=g_i(l)\,Z(\delta l)$
with $Z(\delta l)=s^2\exp(-\lambda^2\delta l/(4\pi v_f))$.
Therefore
\begin{equation}
\frac{\diff}{\diff l}\!\left(\frac{g_i}{g_r}\right)
=\frac{g_r\,\dot{g}_i - g_i\,\dot{g}_r}{g_r^2}
=\frac{(2-K^{-1})g_ig_r-(2-K^{-1})g_rg_i}{g_r^2}=0.
\end{equation}
The ratio $\mathcal{I}=g_i/g_r$ is algebraically conserved to all orders in
the Wilson expansion, not just to leading order.
In spin-boson language, $\mathcal{I}_0\propto\mu/v_f$ is set by the microscopic
bias [Eq.~\eqref{eq:I_from_mu}] and is unchanged by the RG flow.
The EP condition $\mathcal{I}_0=1$ is therefore protected: a system tuned to
$\mu=\mu_c$ remains on the EP under coarse-graining.

\bibliographystyle{apsrev4-2}

\begin{thebibliography}{99}

\bibitem{Berezinskii1971}
V.~L. Berezinskii, Sov.~Phys.~JETP \textbf{32}, 493 (1971).

\bibitem{Kosterlitz1973}
J.~M. Kosterlitz and D.~J. Thouless, J.~Phys.~C \textbf{6}, 1181 (1973).

\bibitem{Kosterlitz1974}
J.~M. Kosterlitz, J.~Phys.~C \textbf{7}, 1046 (1974).

\bibitem{Coleman1975}
S.~Coleman, Phys.~Rev.~D \textbf{11}, 2088 (1975).

\bibitem{Soluyom1979}
J.~S\'olyom, Adv.~Phys.~\textbf{28}, 201 (1979).

\bibitem{Giamarchi2004}
T.~Giamarchi, \textit{Quantum Physics in One Dimension}
(Oxford University Press, Oxford, 2004).

\bibitem{Weiss1999}
U.~Weiss, \textit{Quantum Dissipative Systems}, 2nd ed.\
(World Scientific, Singapore, 1999).

\bibitem{Sieberer2016}
L.~M. Sieberer, M.~Buchhold, and S.~Diehl,
Rep.~Prog.~Phys.~\textbf{79}, 096001 (2016).

\bibitem{Keldysh1965}
L.~V. Keldysh, Sov.~Phys.~JETP \textbf{20}, 1018 (1965).

\bibitem{Kamenev2011}
A.~Kamenev, \textit{Field Theory of Non-Equilibrium Systems}
(Cambridge University Press, Cambridge, 2011).

\bibitem{Mitchell2018}
A.~K. Mitchell and L.~Fritz,
Phys.~Rev.~B \textbf{97}, 115139 (2018).

\bibitem{Sieberer2013}
L.~M. Sieberer, S.~D. Huber, E.~Altman, and S.~Diehl,
Phys.~Rev.~Lett.~\textbf{110}, 195301 (2013).

\bibitem{Bender1998}
C.~M. Bender and S.~Boettcher,
Phys.~Rev.~Lett.~\textbf{80}, 5243 (1998).

\bibitem{Bender2007}
C.~M. Bender, Rep.~Prog.~Phys.~\textbf{70}, 947 (2007).

\bibitem{Bender2013}
C.~M. Bender, M.~Gianfreda, and S.~P. Klevansky,
Phys.~Rev.~A \textbf{88}, 062107 (2013).

\bibitem{Ashida2020}
Y.~Ashida, Z.~Gong, and M.~Ueda,
Adv.~Phys.~\textbf{69}, 249 (2020).

\bibitem{ashida2017parity}
Y.~Ashida, S.~Furukawa, and M.~Ueda,
Phys.~Rev.~A \textbf{95}, 022124 (2017).

\bibitem{PhysRevLett.25.450}
J.~D. Tomonaga, Prog.~Theor.~Phys.~\textbf{5}, 544 (1950).

\bibitem{PhysRevLett.72.892}
J.~M. Luttinger, J.~Math.~Phys.~\textbf{4}, 1154 (1963).

\bibitem{fendley1996unified}
P.~Fendley, A.~W.~Ludwig, and H.~Saleur,
Phys.~Rev.~B \textbf{52}, 8934 (1995).

\bibitem{Jord_wig}
P.~Jordan and E.~Wigner, Z.~Phys.~\textbf{47}, 631 (1928).

\bibitem{Jordan_Wig1}
E.~Lieb, T.~Schultz, and D.~Mattis,
Ann.~Phys.~\textbf{16}, 407 (1961).

\bibitem{malard2013sine}
M.~Malard, H.~Johannesson, and P.~Schlottmann,
Phys.~Rev.~B \textbf{87}, 035120 (2013).

\bibitem{Gaudin1983}
M.~Gaudin, \textit{The Bethe Wavefunction}
(Cambridge University Press, Cambridge, 2014)
[French original: \textit{La Fonction d'Onde de Bethe},
Masson, Paris, 1983].

\bibitem{LiebLiniger1963}
E.~H. Lieb and W.~Liniger,
Phys.~Rev.~\textbf{130}, 1605 (1963).

\bibitem{ZZ1979}
A.~B. Zamolodchikov and A.~B. Zamolodchikov,
Ann.~Phys.~\textbf{120}, 253 (1979).

\end{thebibliography}

\end{document}